\documentclass[12pt]{iopart}   

\usepackage{iopams}  
\usepackage{amssymb}  
\usepackage{graphicx}

\begin{document}  
  
\letter{Gravitational wave detection using electromagnetic 
  modes in a resonance cavity}  
  
\author{Gert Brodin\dag and Mattias Marklund\ddag}  
  
\address{\dag\ Department of Plasma Physics, Ume{\aa} University, 
  SE--901 87 Ume{\aa}, Sweden}  
  
\address{\ddag\ Department of Electromagnetics, Chalmers University 
  of Technology, SE--412 96 G\"oteborg, Sweden}  
  
\begin{abstract}  
We present a proposal for a gravitational wave detector, based on the  
excitation of an electromagnetic mode in a resonance cavity. The mode  
is excited due to the interaction between a large amplitude  
electromagnetic mode and a quasi-monochromatic gravitational  
wave. The minimum metric perturbation needed for detection is estimated  
to the order $7\times 10^{-23}$ using current data on superconducting  
niobium cavities. Using this value together with different standard  
models predicting the occurrence of merging neutron star or black hole  
binaries, the corresponding detection rate is estimated to 1--20
events per year, with a `table top' cavity of a few meters
length.   
\end{abstract}  
  
\submitto{\CQG}

  
\section*{}
During the last decades the quest for detecting gravitational waves has
intensified. The efforts have been inspired by the indirect evidence  
for gravitational radiation \cite{Taylor94}, advances in technology and the  
prospects of obtaining new useful astrophysical information through 
the development of gravitational wave astronomy \cite{gwastronomy}.  
A number of ambitious detector projects are already in operation 
or being built all over the world, for example Ligo and ALLEGRO in USA,  
Virgo, AURIGA and GEO 600 in Europe, Tama 300 in Japan, AIGO and NIOBE  
in Australia \cite{Ligo}. 
Furthermore, there are well developed plans to use space based
gravitational wave detectors, i.e., the LISA 
project \cite{LISA}.  
The detection mechanisms are basically of a mechanical nature in  
the cases above, but there have also been several proposals for  
electromagnetic detection 
mechanisms \cite{Braginsky71}.  
  
In the present paper we will investigate a detection mechanism based
on the interaction of electromagnetic modes and gravitational
radiation in a cavity with highly conducting walls. 
The main feature of our proposed 
gravitational wave detector is that it supports two electromagnetic
eigenmodes with nearby eigenfrequencies, a possibility that has
previously been discussed in Refs.\ \cite{GP,a}. If one 
eigenmode is excited initially (called the pump-mode), and a
quasi-monochromatic gravitational wave with a frequency equal to the
eigenmode frequency difference reaches our system, a new
electromagnetic eigenmode can be excited due to the
gravitational-electromagnetic wave interaction. The coupling mechanism
is similar in principle to the wave interaction processes
described in, e.g., Ref.\ \cite{BM99}.   
  
The low-frequency nature of the gravitational modes, as compared to
the electromagnetic resonance frequencies, at first seem to greatly
limit the efficiency of a cavity with nearby frequencies. To get a
large gravitationally induced mode-coupling  for a simple cavity
geometry, the estimated cavity dimensions become prohibitively large,
i.e., comparable to the wavelength of the gravitational
wave. Solutions to this problem has been found by Refs.\ \cite{a},
who has considered a gravitational wave detector consisting of two
coupled cavities. Cavities based on these principles have been built,
and experimental results are presented in Refs.\ \cite{b}. In
this letter we consider a single cavity with a variable
crossection. The main purpose of varying the crossection is the
following. For a cavity with dimensions much smaller than the
wavelength of the gravitational  wave, the suggested geometry greatly
magnifies the gravitational wave induced mode-coupling.
In our present work we have simulated the effect of a  
varying crossection, by considering a cavity filled with three different  
dielectrics, in order to be able to perform most of the calculations  
analytically. 
It is straightforward to  
make a semi-quantitative translation of our results to the case of a  
vacuum cavity with a varying crossection. Using current  
data on the  
latter type of cavity \cite{Walther2000,Graber,Quality factor}, we
estimate the   
minimum detection level of the metric perturbation to the  
value $h_{\min} \approx 7\times 10^{-23}$, where we have considered an  
inspiraling neutron star or black hole pair as a gravitational wave source.  
If such a level of sensitivity can be reached, neutron  
star or black hole binaries close to collapse could be detected at a  
distance $r_{\max} \sim 10^8 \, \mathrm{ly}$.   
Adopting data from Ref.\ \cite{Event rate} for  
the occurrence of compact binary mergers, we obtain the estimate 
1--20 detection events per year. 

In vacuum, a linearized gravitational wave can 
be represented by $g_{\mu\nu} = \eta_{\mu\nu} + h_{\mu\nu}(\xi)$
where $\xi \equiv x-ct$,  
and $h_{xx} = -h_{yy} \equiv h_{+}, h_{xy} = h_{yx} \equiv h_{\times}$
in standard notation. 

Neglecting terms proportional to derivatives of $h_{+}$  
and $h_{\times}$, the wave equation for the magnetic field is
\cite{APJ,Anile}  
\begin{equation}  
  \left[ \frac{n^2}{c^2}\frac{\partial^2}{\partial t^2}  
  - \nabla^2 \right]{\boldsymbol{B}} = 
  \left[ h_+\left( \frac{\partial^2}{\partial y^2} -  
    \frac{\partial^2}{\partial z^2} \right)  
  + h_{\times}\frac{\partial^2}{\partial y\partial z}  
  \right]{\boldsymbol{B}} ,  \label{WaveB}  
\end{equation}  
and similarly for the electric field. Here $n$ is the index of
refraction. For the moment, we will neglect mechanical
effects, i.e., effects which are associated with the varying
coordinates 
of the walls due to the restoring forces of the cavity.%
\footnote{This is true neglecting thermal noise and assuming a
  long cavity, as compared to the speed of sound over the
  gravitational wave frequency.}

The coupling of two electromagnetic modes and a gravitational wave in a   
cavity will depend strongly on the geometry of the electromagnetic   
eigenfunctions. 
We can greatly magnify the coupling, as compared to
a rectangular prism geometry, by 
varying the cross-section of the cavity, or by filling the cavity partially 
with a dielectric medium. The former case is of more interest 
from a practical point of view, since a vacuum cavity    
implies better detector performance, but we will consider the 
latter case since it can be handled analytically. \emph{However}, we  
will show how to make a semi-quantitative translation of  
our results to the case of a varying cavity cross-section.    
  
Specifically, we choose a rectangular cross-section (side lengths $x_0$ 
and $y_0$), and we divide the length of the cavity into three 
regions. Region 1 has length $l_1$ (occupying the region    
$0<z<l_1$) and a refractive index $n_1$. Region 2 has length $l_2$   
(occupying the region $l_1 < z < l_1 + l_2$), with a refractive index
$n_2$, while region 3 consists of vacuum and has  
length $l_3$ (occupying the region $l_1 + l_2 < z < l_1 +  
l_2 + l_3 \equiv l$). The cavity is supposed to   
have positive coordinates, with one of the corners coinciding with the   
origin. Furthermore, we require that $l_2 \ll l_1$, and that the wave  
number in region 2 is less than in region 1. The reason for this  
arrangement is twofold. Firstly, we want to obtain a large  
coupling between the wave modes, and secondly we want an efficient  
filtering of the eigenmode with the lower frequency in region three.   
  
The first step is to analyze the linear eigenmodes in this system. 
The simplest modes are of the type  
\numparts\label{Region1}  
\begin{eqnarray}  
  E_{y} = \frac{i\omega x_0}{m\pi }\widetilde{B}_{zj}\sin  
\left( \frac{m\pi x}{x_0}%
\right) \sin [k_{j}z + \varphi_{j}]e^{-i\omega t} ,  \\  
  B_{z} = \widetilde{B}_{zj}\cos \left( \frac{m\pi x}{x_0}\right) \sin  
[k_{j}z + \varphi_{j}]e^{-i\omega t} , \\  
  B_{x} = -\frac{k_{j}x_0}{m\pi }\widetilde{B}_{zj}\sin  
\left( \frac{m\pi x}{x_0}%
\right) \cos [k_{j}z + \varphi_{j}]e^{-i\omega t}  , 
\end{eqnarray}  
\endnumparts  
in regions $j=1$, $2$ and $3$, respectively, where the wave in region 3 is a
standing wave, and $m$ is the mode number. 
In region 3 we may also have a decaying wave  
 \numparts\label{Region2b}  
\begin{eqnarray}  
  E_{y} = \frac{i\omega x_0}{m\pi }\widetilde{B}_{z3}\sin \left(
\frac{m\pi x}{x_0}%
\right) \sinh [k_{3}z+\varphi _{3}]e^{-i\omega t} ,  \\  
  B_{z} = \widetilde{B}_{z3}\cos \left( \frac{m\pi x}{x_0}\right) \sinh  
[k_{3}z+\varphi _{3}]e^{-i\omega t} , \\  
  B_{x} = -\frac{k_{3}x_0}{m\pi }\widetilde{B}_{z3}\sin \left( \frac{m\pi
  x}{x_0}%
\right) \cosh [k_{3}z+\varphi _{3}]e^{-i\omega t} .  
\end{eqnarray}  
\endnumparts 
Using standard boundary conditions,   
the wave numbers are  
calculated for an eigenmode, and the relation between the amplitudes in the  
three regions is found, and thereby the mode profile.  
We are interested in the shift from decaying to  
oscillatory behavior in region 3. Denote the highest frequency mode which is
decaying in region 3 with index $a$, and the wave number and decay  
coefficient with $k_{1a}$, $k_{2a}$ and $k_{3a}$
respectively. Similarly, the next   
frequency, which is oscillatory in both regions, is denoted by index
$b$.   
If we have $l \gg x_0$ (and $m$ the same) these two frequencies will
be very close, and a   
gravitational wave which has a frequency equal to the difference between the  
electromagnetic modes causes a small coupling between these modes. An  
example of two such eigenmodes is shown in fig.\ 1.  

\begin{figure}  
\includegraphics[width=\columnwidth]{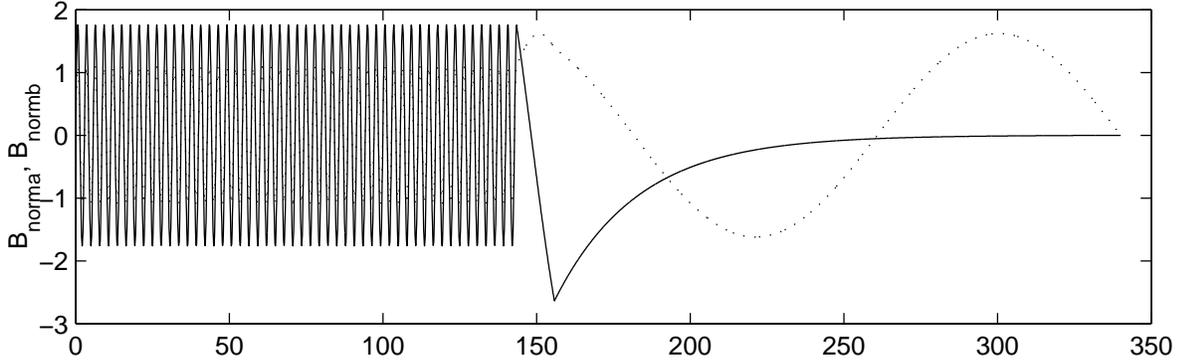} 
\caption{ 
Mode-profiles for the eigenmodes 
are shown for the 
parameter values $n_1= 1.22$, $n_2 = 1.0005$, $l_1m/x_0=143$, 
$l_2m/x_0=12.3$ and $l_3m/x_0=184$. $B_{{\rm norm}\,a}$ is the solid line 
and $B_{{\rm norm}\,b}$ is the dotted line, 
and $I_c = 0.27$.  
} 
\end{figure}

We define the eigenmodes to have the form
$\widetilde{B}_{za,b} = \widetilde{B}_{a,b}(t) 
B_{{\rm norm}a,b}(z,y,z)$, where $\widetilde{B}_{a,b}$ is a   
time-dependent amplitude and the normalized eigenmodes $B_{{\rm norm}a,b}$  
satisfy $\int_{V}\left| B_{{\rm norm}a,b}\right| ^{2}dV=V$.  
We let all electromagnetic field components be of the form
$A = A_a(\boldsymbol{r})\exp(-i\omega_at) +
A_b(\boldsymbol{r})\exp(-i\omega_bt) + \mathrm{c.c.}$, where c.c.\ stands for  
complex conjugate, and the indices stand for the eigenmodes discussed  
above. The gravitational perturbation can be approximated by $h_{+,\times }=%
\widehat{h}_{+,\times }\exp(-i\omega _{g}t) + \mathrm{c.c}$, where we neglect the  
spatial dependency, since the  
gravitational wavelength is assumed to be much longer than all of the cavity  
dimensions. 
During a certain interval in time, the  
frequency matching condition   
  $\omega_b = \omega_a + \omega_g$  
will be approximately fulfilled. Given the wave 
equation (\ref{WaveB}), and the above ansatz we find after integrating 
over the length of the cavity  
\begin{equation}  
\frac{2i\omega _{b}}{c^{2}}\left( \frac{\partial }{\partial t}-\gamma
\right) \widetilde{B}_{b} = -h_{+}k_{1a}^2\widetilde{B}_a %
I_{c} ,  \label{Excitation-eq}  
\end{equation}  
where  
\begin{equation}  
I_{c}=\frac{1}{Vk_{1a}^2}%
\int_{V}\frac{\partial^2B_{{\rm norm}\,a}}{\partial z^2}B_{{\rm
    norm}\,b}\,dV ,  
\label{Coupling-int}  
\end{equation}  
and we have added a phenomenological linear damping term 
represented by  
$\gamma$. Thus we note that for the given geometry, only  
the $h_+$-polarization gives a mode-coupling. 
Calculations of the eigenmode parameters show  
that $I_{c}$ may be different from zero when $n_{1,2}\neq 1$,  
and generally $I_{c}$ of the order of unity can be obtained, see 
fig.\ 1 for an   
example. From Eq.\ (\ref{Excitation-eq}), we find that the saturated  
value of the gravitationally excited mode is  
\begin{equation}  
\left| \widetilde{B}_{b{\rm sat}}\right|
=\frac{h_{+}k_{1a}^2c^2\widetilde{%
B}_{a}}{2\gamma\omega_b}I_c \ .  
\label{Saturation-eq}  
\end{equation}  

In fig.\ 1 it is shown that we can get an appreciable mode-coupling
constant $I_{c}$  
for a cavity filled with materials with different dielectric constants, and  
it is of interest whether or not this can be achieved in a 
vacuum cavity.    
As seen by Eq.\ (\ref{Coupling-int}), the coupling is essentially  
determined by the wave numbers of the modes, given by $k =  
(m^2\pi^2/x_0^2 - n^2\omega^2/c^2)^{1/2}$. 
Thus by adjusting the width $x_0$ in a vacuum cavity,  
we may get the same variations  
in the wave numbers as when varying the index of refraction $n$. 
The translation of our results to a  
vacuum cavity with a varying width is not completely accurate, however.   
When varying $x_0$, the mode-dependence on $x$ and $z$ does not  
exactly factorize, in particular close to the change in width.  
Moreover, the contribution to the coupling $I_c$ in each section becomes  
proportional to the corresponding volume, and thereby also to the  
cross-section.  
However, since most of the contribution to the integral in  
Eq.\ (\ref{Coupling-int}) comes from region 1, our results can  
still be approximately translated  
to the case of a vacuum cavity, by  
varying $x_0$ instead of $n$ such as to get the same wavenumber  
as in our above example.

We denote the minimum detection level of the excited mode with 
$\widetilde{B}_{b\min }$ and the maximum allowed field in the cavity   
with $\widetilde{B}_{a\max }$. 
The magnetic field $\widetilde{B}_{b\min }$ is related to the minimum number
of photons $N_{bm}$ needed for detection by 
$\widetilde{B}_{b\min} \sim \left(N_{bm} \mu_0\omega_b\hbar/x_0y_0l  
\right)^{1/2} $.  Furthermore, we have $2\gamma\omega _{b}/k_{1a}^{2}c^{2}$ 
$\simeq (1/3Q)(\omega_{b}^{2}/k_{1a}^{2}c^2)\simeq 2/3Q$, where 
$Q=\omega _{b}/2\pi\gamma$ is the quality factor of the cavity, and
thus we obtain, using Eq.\ (\ref{Saturation-eq}),     
\begin{equation}\label{hplus} 
  h_{\min} \sim \frac{2}{3QI_c\widetilde{B}_{a\max}}\left( 
  \frac{\mu_0\omega_b\hbar N_{bm}}{x_0y_0l} \right)^{1/2} .
\end{equation} 
Before giving estimates of the parameter values, we will investigate certain other 
effects that may limit the detection efficiency.  

Massive  
binaries produce monochromatic radiation to a good approximation, but close to 
merging, the frequency will be increasing rather rapidly which means 
that the phase matching will be lost. The cavity is designed to detect 
the frequency $\omega_g$, and we have assumed that the signal varies 
as $\exp (-{\rm i}\omega_gt)$, but in reality we have $\exp (-{\rm 
  i}\psi(t) )$, where for simplicity we assume $\psi =0$, $d\psi 
/dt=\omega _{g}$ at $t=0$. 
The coherence
time $t_{{\rm coh}}$ is roughly defined by $\omega _{g}t_{{\rm coh}} -
2\psi(t_{{\rm coh}}/2) = \pi$. Thus, $t_{{\rm coh}} \approx
[2\pi/(d\omega_g/dt)]^{1/2}$ provided $\omega_g t_{\rm coh} \gg 1$.  
Using Newtonian calculations of two 
masses $M$ in circular orbits around the center of mass,  
complemented by the quadrupole formula for gravitational radiation, we  
find  
\begin{equation} 
\frac{d\omega _{g}}{dt}=\frac{388\omega _{g}^{2}}{5c^{5}}\left( \frac{%
GM\omega _{g}}{8}\right)^{5/3} . 
\end{equation} 
Close to binary merging, the time of coherent interaction will be
shorter than the photon life time, and  
in that regime the growth of the excited mode is limited by
decoherence rather than the damping   
due to a finite quality factor of the cavity. However, formula
(\ref{hplus}) can   
still be applied if we simply replace $Q$ by $\omega _{b}t_{{\rm
    coh}}$.

To be able to estimate the number of photons needed for detection, we  
must study various sources of noise. The   
simplest effect is direct thermal excitation of photons in mode 
$b$. Since each mode has a thermal energy level of   
order $KT$, the number of such photons is $\sim KT/\hbar \omega_{b}$.  
However, we will also have a contribution associated with thermal 
variations in the length of the cavity. A 
standard model for the variations in length $\delta l$ is \cite{GP}
\begin{equation}\label{thermalnoise} 
  \frac{d^2 \delta l}{dt^2}+\frac{\omega_l}{Q_{mec}}\frac{d \delta 
  l}{dt}+\omega_l^2 \delta l=- l \omega_g^2 h_+\exp(i \omega_g 
  t)+a_{th}  ,
\end{equation} 
where $\omega_l$ is the eigenfrequency of longitudinal oscillations 
(of the order of the acoustic velocity in the cavity divided by the 
length), $Q_{mec}$ is the mechanical quality factor associated with 
these oscillations, and $a_{th}$ is the stochastic acceleration due 
to the thermal motion, giving rise to a random walk in the 
oscillation amplitude. First we note that the amplitude of the length 
variations with the gravitational frequency is $\delta l_g= \omega_g^2 
h_+/(\omega_g^2-\omega_l^2)$. We will consider the case when 
$\omega_l^2 \sim 2\omega_g^2$, which implies that {\it the amplitude} 
of the gravitational length perturbation is essentially unaffected by 
the restoring force of the cavity, as is the number  
of gravitationally generated photons.  
 
The thermal fluctuation contribution to the right hand side of Eq.\
(\ref{Excitation-eq}), via the  
coupling to the pump wave, becomes proportional to $\delta 
l_{th}\exp[i(\omega_b-\omega_a)t]$. Since $\delta l_{th}\sim 
  \exp(i\omega_l t)$, the longitudinal oscillations give rise 
  to slightly off-resonant (i.e., driven) fields with a frequency 
  difference $\delta \omega=\omega_l-\omega_g$ compared to mode $b$.    
However, due to the stochastic changes in amplitude and/or phase of
the longitudinal oscillation (where the time-scale for significant
changes is given by $\tau\sim Q_{mec}/\omega_l$), a contribution to
mode $b$ of the order
$\left\langle {\delta l_{th}}\right\rangle\sqrt{\omega_l/\delta\omega 
  Q_{mec}}\,dt \sim
\sqrt{KT/m\omega_l^2}\sqrt{1/Q_{mec}}\,(2\pi/\delta\omega)$ 
is also made during a single oscillation period $dt=2\pi/
\delta\omega$. Here $m$ is the mass taking part in the longitudinal
oscillation. During a time of the order $t_{coh}$, this contribution
adds up to approximately
$(\sqrt{2\pi}/\delta\omega)\sqrt{KT/m\omega_l^2}\sqrt{\delta\omega
  t_{coh}/ Q_{mec}}$ by a random walk process. Using
$\delta\omega\sim\omega_g$, the condition for the gravitational
contribution $\sim h l t_{coh}$ to be larger than that of the thermal 
fluctuations can  be written
\begin{equation}\label{h_therm_cond}
h_{min}>\frac{2\pi}{\omega_l
  l}\sqrt{\frac{KT}{m}}\frac{1}{\sqrt{Q_{mec}\omega_g t_{coh}}} .
\end{equation}

Assume that we want to reach a  
sensitivity  $h_{min} \approx 7\times 10^{-23}$. We let
$x_0=0.1 \, \mathrm{m}$, $y_0 = 0.2 \, \mathrm{m}$, $l = 4 \, 
\mathrm{m}$ , and $m=4$ which gives
$\omega_b \approx 4\pi c/x_0 = 4 \times 10^{10} \, \mathrm{rad/s}$.
Furthermore,  
we take $\omega_g = 1000 \, \mathrm{rad/s}$ and let $\omega_l = 1500
\, \mathrm{rad/s}$. We assume  
$T = 1\, \mathrm{K}$ together with $m = 2500\, \mathrm{kg}$. 
From (\ref{h_therm_cond}), we then find that we
need a mechanical quality factor  
$Q_{mec} \approx 6\times 10^7$ to reach the desired
sensitivity.\footnote{Note that mechanical
quality factors can be several orders of magnitude better, see e.g.,
Ref.\ \cite{Braginsky-Rudenko}.} 
Moreover, assuming the necessary number of  
photons for detection to be $N_{bm}=20$ (well above the direct
electromagnetic noise level of a few photons), $I_c \approx
0.3$ and $\widetilde{B}_{a\max} = 0.2 \, \mathrm{T}$
\cite{Graber,Quality factor}, we need an  
electromagnetic quality factor $Q=5\times 10^9$ to reach the desired 
sensitivity   
$h_{\min } \approx  
7\times 10^{-23}$. 
Note that the coherence time is slightly longer than the photon
life-time, as needed.

Following the example given above for calculating the coherence time, 
i.e., a binary consisting of two compact objects, each of one solar 
mass $M_\odot$, separated by a distance $d$, the amplitude of the
metric perturbation at a distance $r$ is given by  
$h \approx 2G^{2}M^{2}/c^2rd$. 
For definiteness we choose $d = 80 \, {\rm km}$ corresponding to  
$\omega_g = 1000 \, \mathrm{rad/s}$, and thus for $h_{\min }$ of the  
order of $7\times 10^{-23} $  
we obtain the maximum observational distance  
  $r_{\max} \approx 8 \times 10^7 \, {\rm ly}$. 
Using data from Ref.\ \cite{Event rate 2}, we  
deduce that the number of galaxies within the observational  
distance $r_{\max} \sim 10^8 \, \mathrm{ly}$  
is of the order $N = 10^{5}$.  
Combining these figures with the expected number of compact binary 
mergers  per galaxy and million  
years \cite{Event rate}, we obtain a detection rate in the interval  
$1-20$ events per year (the uncertainty is due to different 
models used for the birth of compact binaries).

Our proposal for a gravitational wave detector is to a large extent 
based on  currently available 
technology \cite{NOGUES99,Walther2000,Graber,Quality factor}, and   
our requirements are moderate given the performance of
some existing microwave cavities.
For example, values of $Q \approx 4\times 10^{10}$ has been
reported in Ref.\ \cite{Walther2000} and quantum non-demolition
measurements 
of single microwave photons have been made in Ref.\
\cite{NOGUES99}. Furthermore, 
the key performance parameters of the  superconducting niobium
cavities, i.e., 
the quality factor and the maximum  allowed field strength before
field 
emission, have been improving over
the years, suggesting that the detection sensitivity can be increased 
even 
further.
  
A sensitivity $h_{\min} \lesssim 10^{-22}$ seems extremely 
good, but, on the other hand, idealizations have been made when making  
the estimate. 
In addition to any effects induced by the gravitational wave, 
the walls generally vibrate slightly due to the electromagnetic forces  
exerted by the large pump field. While these later oscillations
clearly will be   
larger than the variations in length that are directly due to the  
gravitational wave, the associated nonlinearities will be harmonics of 
the pump frequency, and thus such effects do not couple to the other  
eigenmodes of the cavity. Furthermore, we have assumed that the
detection  
of the excited  mode is not much affected by the presence of the pump 
signals. Even   
though mode $a$ is partially filtered out in region 3, the small  
frequency shift between the two electromagnetic modes may pose a 
certain  difficulty in this respect: In particular, a very narrow
bandwidth (pump) antenna  signal must be used, in order to exclude the 
slightest initial perturbation at the gravitationally excited
frequency $\omega_b$.  

However, although there are technical difficulties in constructing
an electromagnetic detector, the real advantage is the
possibility to reduce the size of the devise. In the
example presented above, the length of the cavity has been taken to be
$4\, \mathrm{m}$, and the cross section roughly $1/50\,
\mathrm{m}^2$. This alone
could prove to be useful when trying to set up new gravitational
wave observatories. 
 
\ack We thank Ulf Jordan for helpful discussions.

\end{document}